\documentclass[prd,preprint,eqsecnum,nofootinbib,amsmath,amssymb,
               tightenlines,dvips]{revtex4}
\usepackage{graphicx}
\usepackage{bm}
\input epsf

\begin {document}


\def\lra{\leftrightarrow}

\def\centerbox#1#2{\centerline{\epsfxsize=#1\textwidth \epsfbox{#2}}}

\def\ca{C_{\rm A}}
\def\cs{C_s}
\def\cf{C_{\rm F}}
\def\ds{d_s}
\def\df{d_{\rm F}}
\def\da{d_{\rm A}}
\def\h{\bm h}
\def\F{{\mathcal F}_s}
\def\Fq{{\mathcal F}_{\rm q}}
\def\Fg{{\mathcal F}_{\rm g}}
\def\Fsol{{\bm F}_s}
\def\ptot{p^{\prime}}
\def\qperp{{\bm q}_\bot}

\def\meffg{m^2_{\rm eff,g}}
\def\meffs{m^2_{{\rm eff},s}}
\def\gammaq{\gamma_{q{\lra}qg}}
\def\gammaqbar{\gamma_{\bar q{\lra}{\bar q}g}}
\def\gammag{\gamma_{g{\lra}q {\bar q}}}
\def\gammagg{\gamma_{g{\lra}gg}}
\def\md{m_{\rm D}}
\def\mdsq{m_{\rm D}^2}

\def\S{{\rm \bf S}}
\def\Real{{\rm Re}}

\def\alphas{\alpha_{\rm s}}
\def\p{{\bm p}}
\def\k{{\bm k}}
\def\QT{Q_\perp}
\def\QThat{\hat Q_\perp}
\def\QTo{\hat Q_{\perp0}}
\def\gammaE{\gamma_{\rm E}^{}}
\def\Nf{N_{\rm f}}

\def\q{{\bm q}}
\def\ta{t_{\rm A}}
\def\tf{t_{\rm F}}

\def\B{{\bm B}}
\def\j{{\bm j}}
\def\O{O}

\def\bell{{\bm\ell}}



\title
    {
     QCD Splitting/Joining Functions at Finite Temperature
     in the Deep LPM Regime
    }

\author{Peter Arnold and \c{C}a\=glar Do\=gan}
\affiliation
    {%
    Department of Physics,
    University of Virginia, Box 400714,
    Charlottesville, Virginia 22904, USA
    }%

\date {\today}

\begin {abstract}%
    {%
      There exist full leading-order-in-$\alphas$ numerical calculations of the
      rates for massless quarks and gluons to split and join
      in the background of a quark-gluon plasma through
      hard, nearly collinear bremsstrahlung and inverse
      bremsstrahlung.  In the limit of partons with very high energy $E$,
      where the physics is dominated by the Landau-Pomeranchuk-Migdal
      (LPM) effect,
      there are also analytic leading-log calculations of these rates,
      where the logarithm is $\ln(E/T)$.
      We extend those analytic calculations to next-to-leading-log order.
      We find agreement with the full result to within roughly 20\%
      for $E_< \gtrsim 10 T$, where $E_<$ is the energy of the
      least energetic parton in the splitting/joining process.
      We also discuss how to account for the running of the coupling
      constant in the case that $E/T$ is very large.
      Our results are also applicable to isotropic non-equilibrium plasmas
      if the plasma does not change significantly over the formation time
      associated with particle splitting.
}%
\end {abstract}

\maketitle
\thispagestyle {empty}


\section {Introduction and Results}
\label{sec:intro}

When very high energy particles travel through a quark-gluon or
electromagnetic plasma, the dominant energy loss mechanism is through
hard bremsstrahlung or pair creation, similar to the cascading of high
energy cosmic rays in the atmosphere or of a high energy
particle in a calorimeter.
It was long a problem of interest to calculate the rate for such
splitting processes in the formal limit of very high temperature
quark-gluon plasmas, where the running strong coupling $\alphas(T)$ can
be treated as small
\cite{Zakharov,BDMPS1,BDMPS2,Betc,BSZreview,JeonMoore}.
The problem is complicated by the
Landau-Pomeranchuk-Migdal (LPM) effect \cite{Migdal}
(reviewed below):
for very high energy particles, the quantum mechanical
duration of the splitting process exceeds the mean free time between
collisions, and so successive collisions cannot be treated independently.

For the case of approximately on-shell massless particles traveling
through an infinite medium, a complete leading-order analysis of such
processes was carried out by Jeon and Moore \cite{JeonMoore}
using the formalism
of Arnold, Moore, and Yaffe (AMY) \cite{AMY1,AMY2,AMY3}.
This analysis requires
substantial numerical work to solve integral equations describing the LPM
effect.  Where possible, it's always nice to have analytic results in
place of numerical results, because they are simpler to calculate and
because they can facilitate comparison between different approaches.
One case where analytic results can be found,
explored in earlier literature, is the limit where the
particle momentum $p$ is so large compared to the plasma temperature $T$
that the inverse logarithm $1/\ln(p/T)$ can be treated as a small
number.
We will refer to this as the deep LPM regime.
In the limit of small $\alphas$, earlier authors
have given analytic results for splitting processes to leading order in
powers of this inverse logarithm.%
\footnote{
   See Eqs.\ (19--20) of Ref.\ \cite{bottom_up}, which is based on
   the earlier work of Refs.\ \cite{Zakharov,Betc,BSZreview}.
}
(In contrast, the work of Jeon and Moore made no
assumption about the size of the logarithm, and holds both in and out of
the deep LPM limit, to leading order in coupling $\alphas$.)
The difficulty with leading-log results is that for practical
purposes there is a huge difference, for example,
between $\ln(p/T)$ and $\ln(p/4\pi^2 T)$ if $p$ is of order 10--100 $T$.
But a leading-log analysis will not distinguish between these two
situations since $\ln(p/4\pi^2 T) = \ln(p/T) + O(1)$.
Consider, for example, 5--20 GeV jets at early times in a RHIC
collision, with temperatures of order 300 MeV.
As a general rule, one almost always
needs to push expansions in inverse logarithms
to next-to-leading-log order to get useful results.

Our goal is to find analytic results for splitting processes to
next-to-leading logarithmic (NLL) order in the deep LPM regime.  In this
paper, we make a first step towards that goal by computing the
next-to-leading logarithmic result in the formal parametric
limit that $T \ll p \ll T/[\alphas^2 \ln(\alphas^{-1})]$ for weak coupling:
that is, deep in the LPM regime, but not too deep.
We will leave the more involved calculation of NLL results
for higher energy
particles $p \gtrsim T/[\alphas^2 \ln(\alphas^{-1})]$ for future work.
However, later in this paper (section \ref{sec:largeA}),
we will see that NLL results for
higher-energy particles will differ by only about 15\% from the
formulas derived here.
In this paper we will also discuss
leading-log results in the
case of extremely high energy particles, where $P$ is so large
that there is significant difference between the running
couplings $\alphas(P)$ and $\alphas(T)$.

As we will discuss later, our results apply not just to
equilibrium plasmas but more generally to plasmas with an isotropic
distribution of particle momenta (in which case the ``$T$'' in our
discussion refers to the typical energy scale of plasma particles).  We
express our answers in terms of the splitting functions $\gamma_{a
\leftrightarrow bc}\bigl(P; xP, (1-x)P\bigr)$ of Refs.\
\cite{AMY2,AMY3},%
\footnote{
  We use the term ``splitting functions'' to describe the
  functions $\gamma$ in the splitting rate (\ref{eq:dGamma}).
  Though there is some connection, they should not be confused with
  the DGLAP splitting functions familiar from zero-temperature
  QCD.
  Also, in the notation of Refs.\ \cite{AMY2,AMY3}, this would be
  $\gamma^a_{bc}\bigl(P; xP, (1-x)P\bigr)$.
}
which are
defined so that the rate per particle of type $i$ and momentum $P$
for that particle to split
into particles of type $j$ and $k$ and momentum fractions $x$ and
$1-x$ is%
\footnote{
  Readers of Ref.\ \cite{JeonMoore} should note that we use the symbol
  $\Gamma$ to denote rate per particle, whereas Ref.\ \cite{JeonMoore}
  denotes this by
  $d\Gamma/dt$.
}
\begin {equation}
  \frac{d\Gamma_{a\to bc}}{dx}
  = \frac{(2\pi)^3}{P\nu_a} \,
    \gamma_{a \leftrightarrow bc}\bigl(P; xP, (1-x)P\bigr) \,
    [1\pm f_b(xP)] \, [1\pm f_c((1-x)P)] .
\label {eq:dGamma}
\end {equation}
Here $f(p)$ is the phase space distribution of plasma particles of
a given type,
and the factors $1\pm f(p)$ are
Bose enhancement or Fermi blocking factors.
In equilibrium,
$f(p) = 1/(e^{\beta p} \mp 1)$ is the
Bose or Fermi distribution
associated with (massless) particles of a given type.
$\nu_a$ is the number of spin times color states for particle type $a$
({\it i.e.}\/ 6 for a quark or anti-quark and 16 for a gluon).
The prefactors in (\ref{eq:dGamma}) are just convenient
normalization conventions
in the definition of $\gamma_{a \leftrightarrow bc}$.

\begin{figure}
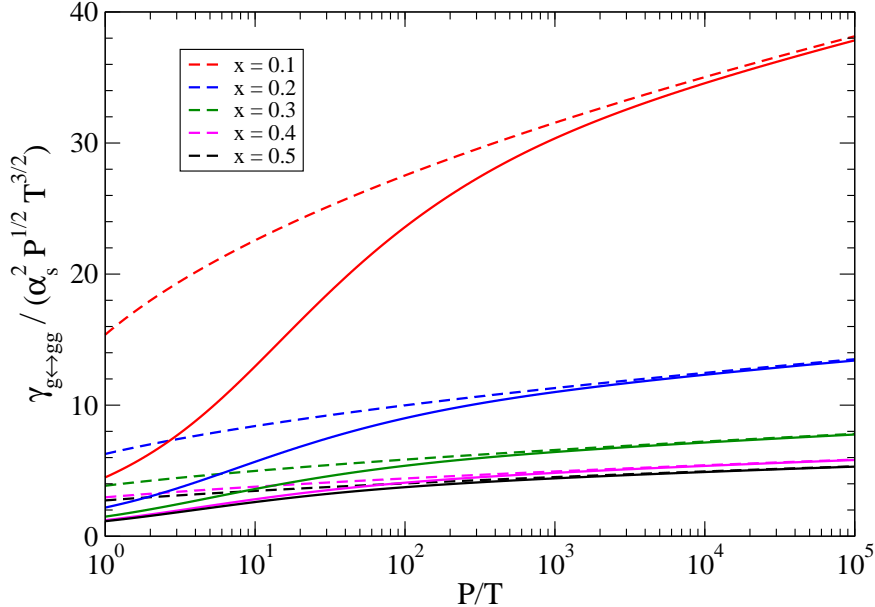

  \centerbox{0.7}{ggg_Nf3_p.eps}
  \caption{\label{fig:ggg}
  Exact (solid line) and next-to-leading logarithmic order (dashed line)
  results for
  $\gamma_{g \leftrightarrow gg}$
  at leading order in coupling $\alphas$
  for QCD with three massless flavors ($N_f=3$).
  $\gamma_{g \leftrightarrow gg}$
  is plotted in units of $\alphas^2 P^{1/2}T^{3/2}$ as a function
  of $P/T$ for various values of $x$,
  where $P$ is the momentum of the initial high-energy particle
  and $xP$ and $(1-x)P$ are the momenta after splitting.
  From top to bottom, the curves corresponds to
  $x = 0.1$, 0.2, 0.3, 0.4, 0.5.
}
\end{figure}

We will give the formula for our next-to-leading-log (NLL) result in Sec.\
\ref{sec:result}, after a qualitative review of the form of
the leading-log result.  For now, we offer
in Fig.\ \ref{fig:ggg}
a numerical comparison of
the NLL computation to the full leading-order-in-$\alphas$ formula
of Refs.\ \cite{AMY2,AMY3}
for the case of gluon splitting $g\leftrightarrow gg$.
Formally, both of these curves assume weak coupling and that
$P$ is parametrically small compared to $T/[\alphas^2 \ln(\alphas^{-1})]$.
Obviously, for any phenomenologically interesting values of $\alphas$,
the last assumption is rather unlikely to be valid for the momenta
$P \sim 10^5 \, T$
shown at the far right of the plot.  We extend the plot this far
simply to show that the NLL curve is successfully approaching the
more complete numerical calculation based on the same assumption,
as it must.

\begin{figure}
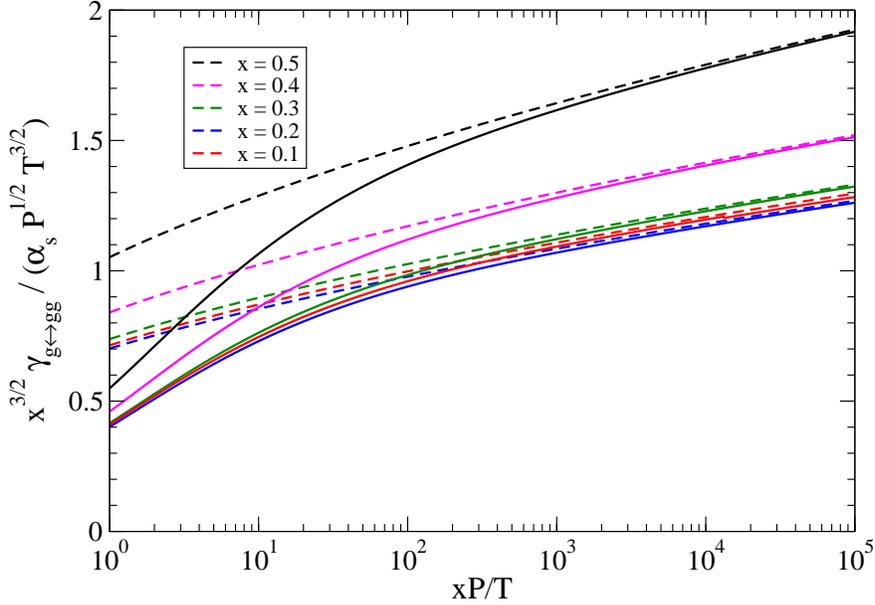

  \centerbox{0.7}{ggg_Nf3_p_rescale.eps}
  \caption{\label{fig:ggg_rescale}
  As Fig. \ref{fig:ggg} but the horizontal axis shows $p_</T$,
  which is $xP/T$ for the $x$ values used to label
  the curves, and the vertical axis has been rescaled by a factor
  of $x^{3/2}$.
  From top to bottom, the curves corresponds to
  $x = 0.5$, 0.4, 0.3, 0.1, 0.2.
}
\end{figure}

To summarize how well the NLL expansion works,
it is useful to rescale the plot as in Fig.\ \ref{fig:ggg_rescale}.
Here, the horizontal axis shows the smallest of the two final
momenta,
\begin {equation}
   p_< \equiv \min\bigl(xP,\, (1-x)P\bigr) ,
\end {equation}
instead of the initial
particle momentum $P$, and the vertical axis has been scaled with
$x$ in a way that shows the limiting small $x$ behavior by
collapsing small $x$ curves atop each other.
For $p_< \simeq 100T$, the NLL result differs
from the full leading-order formula by roughly 5\%.
For $p_< \simeq 10T$, the difference is roughly 20\%.
Pushing the expansion in inverse logarithms
down to $p_</T \simeq 1$,
where one would not expect it to be useful, we see that
the NLL result gives the right order of magnitude
but is off by a factor of roughly 2.
We conclude that the NLL approximation for gluon splitting
in an infinite medium is reasonable for $p_< \gtrsim 10 T$.

In the next section, we will set up our discussion by giving a brief
qualitative review of the basic parametric scales associated with the
LPM effect in a quark-gluon plasma.
We also discuss the reason for the parametric assumption
$P \ll T/[\alphas^2 \ln(\alphas^{-1})]$
in our analysis: In this limit, it turns out
that the momentum transfer due to scattering during a splitting
process is small compared to the $O(T)$ momenta of typical plasma
particles, which greatly simplifies the analysis.
In section \ref{sec:result}, we
present the formulas for our NLL result, followed by their derivation in
section \ref{sec:details}.  In section
\ref{sec:numerics}, we compare numerical results for
processes besides the $g \to gg$ splitting presented above.

We should clarify that throughout most of this paper, we work to
leading order in powers of the coupling $\alphas$.  In particular,
the phrase ``leading logarithm'' will mean the leading logarithmic
contribution at leading order in $\alphas$.  It does not mean a
sum of leading logarithms at all orders in $\alphas$.  The
phrase next-to-leading logarithm will be used similarly, and does
not include any resummation of effects higher order in $\alphas$.

In particular, our NLL results will formally assume that $\ln(P/T)$ is
parametrically large but that $\alphas \ln(P/T)$ is parametrically
small.  In section \ref{sec:run}, we discuss what to do when
$\alphas \ln(P/T)$ is not small.  This is equivalent to a discussion of what
renormalization scale should be used when evaluating $\alphas$.
In this section, we necessarily abandon the
$P \ll T/[\alphas^2 \ln(\alphas^{-1})]$
restriction,
considering much higher momentum scales,
but we will only work to leading log order.


\section{Qualitative Review of Leading-Log Result}
\label {sec:qualitative}

\subsection {Basic Parametric Estimates}

\begin{figure}
  \centerbox{0.3}{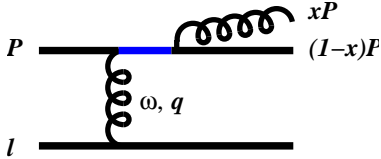}
  \caption{\label{fig:diagram}
  A diagram contributing to bremsstrahlung from a single collision.
}
\end{figure}

Before presenting our next-to-leading-log results, it is
useful to first qualitatively review the form of leading-log results.
For definiteness, we'll consider the case of gluon
bremsstrahlung.  Splitting can occur when nearly
collinear bremsstrahlung accompanies a small-angle scattering
of the high energy particle off of a plasma particle, such as depicted
by the diagram of Fig.\ \ref{fig:diagram} for a momentum $P$ particle
splitting into two particles of momentum $k \simeq xP$ and $p \simeq (1-x)P$.
The intermediate solid
line in this diagram is off-shell in energy by an amount of order
\begin {align}
   \delta E
   &\equiv E_\p + E_\k - E_{\p+\k}
\nonumber\\
   &\simeq
   \frac{p_\perp^2 + m^2}{2p}
   + \frac{k_\perp^2 + m_{\rm g}^2}{2k}
   - \frac{|\p_\perp + \k_\perp|^2 + m^2}{2(p+k)}
\nonumber\\
   &\simeq
   \frac{p_\perp^2 + m^2}{2xP}
   + \frac{k_\perp^2 + m_{\rm g}^2}{2(1-x)P}
   - \frac{|\p_\perp + \k_\perp|^2 + m^2}{2P} \,.
\end {align}
For simplicity, in this review section we'll focus on the
case $x \sim 0.5$ (that is, neither $x$ nor $1{-}x$ very small).
In a typical bremsstrahlung process, the relative angle between
$\p$ and $\k$ is the same order of magnitude as
the angle of deflection in the initiating small-angle
scattering process.%
\footnote{%
\label {foot:angles}%
   For larger angles between $\p$ and $\k$, there is a cancellation
   between the amplitudes for initial and final state radiation.
   In vacuum, bremsstrahlung can be logarithmically
   dominated by smaller angles
   between $\p$ and $\k$, giving rise to a collinear logarithmic
   enhancement of the bremsstrahlung rate.
   This is nor relevant to the deep LPM limit, however, for
   reasons that will be reviewed in footnote \ref{foot:logs}.
}
For $x \sim 0.5$, then
\begin {equation}
  \delta E \sim \frac{\QT^2 + (\mbox{masses})^2}{P} \,,
\label {eq:deltaE1}
\end {equation}
where $\QT$ is the size of the momentum transfer (transverse to the
high energy particle's direction of motion) in the underlying scattering
process.  Thermal particle masses are of order $gT$, and the
relevant range of $\QT$ is bounded below by the
Debye screening mass of order $gT$.  So we can simplify the
parametric estimate (\ref{eq:deltaE1}) to
\begin {equation}
  \delta E \sim \frac{\QT^2}{P} \,.
\end {equation}
The quantum duration $1/\delta E$ of the Bremsstrahlung process is
called the ``formation'' time of the Bremsstrahlung gluon:
\begin {equation}
  t_{\rm form} \sim \frac{1}{\delta E} \sim \frac{P}{\QT^2} \,.
\label {eq:tform}
\end {equation}
In the limit of large $P$, this time will become larger than the
mean free time between collisions, and successive collisions can no
longer be treated as quantum mechanically independent for the
calculation of bremsstrahlung.  As far as bremsstrahlung is concerned,
the multiple collisions during the formation time have roughly the same
effect as a single collision with the same total momentum transfer
$\QT$ as the multiple collisions.  So, on the one hand, the
formation time is given by (\ref{eq:tform}).  On the other hand,
it is also the time $t_{\rm coll}$
for multiple collisions to generate a
total momentum transfer of size $\QT$, which is given by%
\footnote{
  For a review of these scales in extremely simple language, see
  Sec.\ 4.5.1 of Ref.\ \cite{brazil} in the context of $P \sim T$.
  For a more serious discussion in the deep LPM regime, see
  for example Sec.\ 4.1 of Ref.\ \cite{BDMPS1}, where the translation
  to the present discussion is that their mean free path
  $\lambda$ for $2{\to}2$ scattering is of order $1/(g^2 T)$
  and their inverse screening length $\mu$ is of order $gT$.
}
\begin {equation}
   t_{\rm coll}^{-1}
   \sim n \, \sigma_1(Q_\perp) \, \ln\left(\frac{\QT}{\md}\right)
   \sim \frac{g^4 T^3}{\QT^2} \, \ln\left(\frac{\QT}{\md}\right) .
\label {eq:tcoll}
\end {equation}
Here, $n \sim T^3$ is the density of particles to scatter from,
and $\sigma_1(\QT) \sim g^4/\QT^2$ is the cross-section for
a {\it single} Coulomb-like scattering with momentum transfer of order $\QT$.
The logarithmic enhancement to the rate $t_{\rm coll}^{-1}$
is known as a Coulomb logarithm, which accounts for the fact that a
given total momentum transfer can occur not only through a single collision,
but also through multiple collisions
which each individually transfer less momentum but occur more
frequently.%
\footnote{
  For a textbook discussion of Coulomb logarithms, see
  Ref.\ \cite{kinetic}.
}
Self-consistently equating (\ref{eq:tform}) and (\ref{eq:tcoll}),
\begin {equation}
  \frac{P}{\QT^2}
  \sim t_{\rm form} \sim t_{\rm coll}
  \sim \frac{\QT^2}{g^4 T^3 \ln(\QT/\md)} .
\label {eq:tconsistent}
\end {equation}
This determines the total $\QT$ and thence $t_{\rm form}$ and
$t_{\rm coll}$.  Finally, we want the rate for bremsstrahlung.
Every time increment $t_{\rm coll}$, we produce effectively one net collision
from the point of view of a potential bremsstrahlung gluon.  The
cost of adding a gluon emission to a scattering process is
parametrically a factor%
\footnote{%
\label {foot:logs}%
   For bremsstrahlung from a single collision in vacuum,
   there is also generally both an infrared
   logarithmic enhancement, associated with integrating over small $x$,
   and a collinear logarithmic enhancement, previously mentioned
   in footnote \ref{foot:angles}.
   The infrared behavior need not be considered here because
   we are restricting attention to $x \sim 0.5$ in this qualitative
   discussion.  (Note also that the discussion of the infrared in medium
   would be different from the vacuum case
   because of the final state factor $1+f$ for the bremsstrahlung
   particle.)
   The collinear logarithm in the vacuum case depends on the initial
   and final state particles traveling in straight lines for
   sufficiently long
   distances before and after the collision.
   It does not arise in the LPM regime because
   $t_{\rm coll} \sim t_{\rm form}$.
   In any case, we will shortly turn to specific results rather
   than rough parametric estimates, and the only logarithm that
   one finds in the answer is the Coulomb logarithm in
   (\ref{eq:tcoll}).
}
of $g^2$.  So the rate for bremsstrahlung is of order $g^2/t_{\rm coll}$,
except that we need to include final state factors $1 \pm f$:
\begin {equation}
  \Gamma_{\rm brem}
  \sim \frac{g^2}{t_{\rm coll}}
       \times \mbox{(final state factors)} .
\end {equation}
Our qualitative discussion in this section has been for $x \sim 0.5$,
and the $\Gamma_{\rm brem}$ above is implicitly integrated over $x$'s
of this order.
Comparing to (\ref{eq:dGamma}), the splitting functions that
we defined earlier are then of order
\begin {equation}
  \gamma \sim \frac{g^2 P}{t_{\rm coll}}
  \sim g^2 \QT^2 .
\label {eq:gammatilde}
\end {equation}
We can now determine $\QT$ self-consistently from
(\ref{eq:tconsistent}) and thence
the splitting function via (\ref{eq:gammatilde}).


\subsection {Preview of the inverse log expansion}
\label {sec:iteration}

To make the leading-log expansion more explicit, it's helpful to note
that the Debye mass $\md$ is of order $gT$ and rewrite
the parametric equation (\ref{eq:tconsistent}) for $\QT$ in the form
\begin {equation}
   \QThat^2 \sim
   \left(\frac{P}{T}\right)^{1/2}
   \ln^{1/2}(\QThat^2) ,
\label {eq:QT}
\end {equation}
where
\begin {equation}
   \QThat \equiv \frac{\QT}{\md} \,,
\end {equation}
and then expand the solution for $\QThat$ by iteration.
First ignore the logarithm and define $\QTo$ by the corresponding solution
\begin {equation}
   \QTo^2 \sim \left(\frac{P}{T}\right)^{1/2} .
\label {eq:QTo}
\end {equation}
Plugging this into the right-hand side of (\ref{eq:QT}) gives
an improved solution
\begin {equation}
   \QThat^2
   \sim \left(\frac{P}{T}\right)^{1/2} \ln^{1/2}(\QTo^2)
\label {eq:Qperp1}
\end {equation}
So
\begin {equation}
   \QThat^2
   \sim \left(\frac{P}{T}\right)^{1/2} \ln^{1/2}\left(\frac{P}{T}\right) ,
\label {eq:Qperp2}
\end {equation}
and the splitting function (\ref{eq:gammatilde}) is
\begin {equation}
  \gamma
  \sim g^2 \md^2 \QThat^2
  \sim g^4 T^2 \left(\frac{P}{T}\right)^{1/2}
       \ln^{1/2}\left(\frac{P}{T}\right) .
\label {eq:gammatilde2}
\end {equation}

The inverse logarithm expansion is predicated on the assumption that
the logarithm in (\ref{eq:Qperp1}) and (\ref{eq:Qperp2})
is large.  As an example, supposed
we plug the approximation (\ref{eq:Qperp1}) back into the right-hand
side of (\ref{eq:QT}) to get a new approximation for $\QThat$:
\begin {align}
   \QThat^2
   &\sim \left(\frac{P}{T}\right)^{1/2}
    \ln^{1/2}\left[\left(\frac{P}{T}\right)^{1/2} \ln^{1/2}(\QTo^2)\right]
\nonumber\\
   &= \left(\frac{P}{T}\right)^{1/2}
      \left[ \ln\left(\frac{P}{T}\right)
       + \ln \ln (\QTo^2) \right]^{1/2} .
\end {align}
In the limit that $\ln(P/T) \gg 1$, then $\ln(P/T)$ is also large
compared to $\ln\ln(\QTo^2) \sim \ln\ln(P/T)$.  So we may formally choose to
think of the double-log correction as part of the next-to-leading order
terms in the expansion in inverse logs.


\subsection {The parametric assumption
  \boldmath$P \ll T/[\alphas^2\ln(\alphas^{-1})]$}

As we'll discussion in section \ref{sec:eqs}, the NLL calculation will
be much simpler if the momentum transfers in individual scattering
events can be treated as a small perturbation to typical particles in
the plasmas.  The relevant transverse momentum transfers $q_\perp$
of individual collisions range from order $\md$ to order $Q_\perp$,
which gives rise to the Coulomb logarithm in (\ref{eq:tcoll}).
So the simplifying assumption that $q_\perp \ll T$ is equivalent
to $Q_\perp \ll T$ in our analysis.  From (\ref{eq:QT}), this is
\begin {equation}
  \frac{T^2}{\md^2}
  \gg \left(\frac{P}{T}\right)^{1/2} \ln^{1/2}\left(\frac{T^2}{\md^2}\right)
  ,
\end {equation}
which in turn is
\begin {equation}
  P \ll \frac{T^5}{\md^4 \ln(T^2/\md^2)}
    \sim \frac{T}{g^4 \ln(1/g)} .
\end {equation}
In weak coupling, the simplifying assumption that momentum transfers
are small compared to $T$ is therefore parametrically
$P \ll T/[\alphas^2\ln(\alphas^{-1})]$.


\section {Our NLL result}
\label {sec:result}

Our next-to-leading-logarithm result can be summarized in the following
form:
\begin {subequations}
\label{eq:gammas}
\begin {align}
   \gamma_{g\leftrightarrow gg}(P; xP, (1-x)P)
   &= \frac{\da \ca \alphas}{(2\pi)^4 \sqrt2}
     \, \md^2 \, \hat\mu_\perp^2(1,x,1{-}x;{\rm A},{\rm A},{\rm A}) \,
     \frac{1+x^4+(1-x)^4}{x^2(1-x)^2} \,,
\label {eq:gamma_ggg}
\\
   \gamma_{q\leftrightarrow gq}(P; xP, (1-x)P)
   &= \frac{\df \cf \alphas}{(2\pi)^4 \sqrt2}
     \, \md^2 \, \hat\mu_\perp^2(1,x,1{-}x;{\rm F},{\rm A},{\rm F}) \,
     \frac{1+(1-x)^2}{x^2(1-x)} \,,
\\
   \gamma_{g\leftrightarrow q\bar q}(P; xP, (1-x)P)
   &= \frac{\df \cf \alphas}{(2\pi)^4 \sqrt2}
     \, \md^2 \, \hat\mu_\perp^2(1,x,1{-}x;{\rm A},{\rm F},{\rm F}) \,
     \frac{x^2+(1-x)^2}{x(1-x)}
\label{eq:gamma_gqq}
\\ &
   \hspace{12em}
   \mbox{(per quark flavor)} ,
\nonumber
\end {align}
\end {subequations}
where $\hat\mu_\perp^2(x_1,x_2,x_3;s_1,s_2,s_3)$ solves the equation
\begin {align}
  \hat\mu_\perp^2 &=
  \frac{gT}{\md}
  \left[
    \frac{2}{\pi} \,
    x_1 x_2 x_3 \,
    \frac{P}{T}
  \right]^{1/2}
  \Biggl[
       \tfrac12 (C_{s_2}+C_{s_3}-C_{s_1}) x_1^2
               \ln\left(\frac{\xi\hat\mu_\perp^2}{x_1^2}\right)
\nonumber\\ & \qquad \qquad
       + \tfrac12 (C_{s_3}+C_{s_1}-C_{s_2}) x_2^2
               \ln\left(\frac{\xi\hat\mu_\perp^2}{x_2^2}\right)
       + \tfrac12 (C_{s_1}+C_{s_2}-C_{s_3}) x_3^2
               \ln\left(\frac{\xi\hat\mu_\perp^2}{x_3^2}\right)
    \Biggr]^{1/2} ,
\label {eq:muhat}
\end {align}
$\xi$ is the constant
\begin {equation}
   \xi \equiv \exp\left(2 - \gammaE + \tfrac{\pi}{4}\right) 
   \simeq 9.09916 \,,
\label {eq:xi}
\end {equation}
$\gammaE$ is the Euler-Mascheroni constant, and the other constants will
be defined in a moment.  Schematically, these equations are of the form
(\ref{eq:QT}) and (\ref{eq:gammatilde2}) with $\hat\mu_\perp$ playing
the role of $\QThat$.  The appearance of three different logarithms can
be roughly understood as arising because there are three different
particle momenta in the problem: $P$, $xP$, and $(1-x)P$.  (In contrast,
the earlier qualitative discussion was only about orders of magnitude
and took $x \sim 0.5$.)  The various explicit functions of $x$ at the
ends of Eqs.\ (\ref{eq:gammas}) are simply the usual vacuum DGLAP
splitting functions divided by $x(1-x)$.  The role of the medium,
both in terms of providing the momentum transfer and the LPM
effect, is contained in $\md^2 \hat\mu_\perp^2$.

The reason we have switched notation from $\QThat$ to $\hat\mu_\perp$
is because there are several different transverse momenta in the
problem, associated with the three different particles, and each
of them has a distribution of values rather than a single
well-defined value.  We did not want to give the impression that
$\hat\mu_\perp$ exactly corresponded to a particular transverse momentum
in the problem.  For $x \sim 0.5$, $\hat\mu_\perp$ is the same order
of magnitude as the $\QThat$ scale we identified in earlier discussion,
but one could just as well redefine the normalization of
$\hat\mu_\perp$ by replacing the symbol $\hat\mu_\perp$ by
$c\hat\mu_\perp$ everywhere on both sides of equations (\ref{eq:gammas}) and
(\ref{eq:muhat}), for some numerical constant $c$.

In the preceding equations,
$\cf$ and $\ca$ are the quadratic Casimirs, and $\df$ and $\da$
are the dimensions, of the
fundamental and adjoint color
representations.
It's also convenient to define the trace normalization factor $t_R$
by $\operatorname{tr}(T_R^a T_R^b) = t_R \delta^{ab}$, where $T_R^a$ are color
generators.  In general, $t_R = d_R C_R/ d_A$ and $\ta = \ca$.
For QCD,
\begin {equation}
   \cf = \tfrac43 \,,
   \qquad
   \ca = 3 ,
   \qquad
   \df = 3 ,
   \qquad
   \da = 8 ,
   \qquad
   \tf = \tfrac12 ,
   \qquad
   \ta = 3.
\end {equation}
For QCD with $\Nf$ massless fermion flavors, the Debye mass is
given by
\begin {equation}
   \md^2 = \Bigl( \ta + \Nf \tf \Bigr)
           \tfrac{1}{3} g^2 T^2
   = \left( 1 + \tfrac16 \Nf \right) g^2 T^2 .
\label {eq:md}
\end {equation}
Also, the $\gamma_{g \leftrightarrow q\bar q}$ formula in
(\ref{eq:gamma_gqq}) is for a single flavor of quark in the final
state, and so (\ref{eq:dGamma}) should be multiplied by a factor
of $\Nf$ if one wants the total rate for a gluon to split into any
$q\bar q$ pair.

One can formally expand in powers of inverse logarithms by
solving (\ref{eq:muhat}) by iteration, as discussed qualitatively
in section Sec.\ \ref{sec:iteration}.  If some
initial guess $\QTo \sim (P/T)^{1/2}$ is made for $\hat\mu_\perp$,
then the iterated approximations are
\begin {align}
   \hat\mu_{\perp1}^2 &= R(\QTo^2) ,
\\
   \hat\mu_{\perp2}^2 &= R(\hat\mu_{\perp1}^2)
       =R(R(\QTo^2)) ,
\label {eq:mu2}
\end {align}
where $R(\hat\mu_\perp^2)$ represents the right-hand side of
(\ref{eq:muhat}).  The result $\hat\mu_{\perp 2}$ is valid
to next-to-leading-log order and is actually the form we derive
our result in later in this paper.
If one changes the initial guess $\QTo$ by an $O(1)$ multiplicative
factor, it only affects the result for $\hat\mu_{\perp 2}$
at yet higher order in the inverse logarithm expansion.
That's fine as a theoretical statement, but it leaves
ambiguous how to choose $\QTo$ for numerical
comparisons such as Figs.\ \ref{fig:ggg} and \ref{fig:ggg_rescale}.
We therefore re-organized our result into the
natural form (\ref{eq:muhat}), which provides a specific prescription
for determining $\hat\mu_\perp$.  The slight cost
is that, instead of a closed form expression like
(\ref{eq:mu2}) for the NLL result,
we have a simple implicit algebraic equation (\ref{eq:muhat})
for $\hat\mu_\perp$
that needs to be solved numerically.


\section {Details of the Calculation}
\label{sec:details}

\subsection {The equations to solve}
\label {sec:eqs}

We will work in the formalism of Ref.\ \cite{AMY2,AMY3}.  In their notation,
the functions $\gamma_{a{\lra}bc}$ describing nearly collinear
splitting/joining are
\begin{subequations}
\label{eq:gamma}
\begin{eqnarray}
  \gammaq(\bar p';\bar p, \bar k) = \gammaqbar(\bar p';\bar p, \bar k)
  & = & \frac{\bar p'^2+\bar p^2}{\bar p'^2 \, \bar p^2 \, \bar k^3} \,
        \Fq(\bar p', \bar p, \bar k) \, ,
\\
\label {eq:gammagqq}
  \gammag(\bar p'; \bar p, \bar k)
  & = & \frac{\bar k^2+\bar p^2}{\bar k^2 \, \bar p^2 \, \bar p'^3}
        \, \Fq(\bar k, -\bar p,\bar p') \, ,
\\
  \gammagg(\bar p'; \bar p, \bar k)
  & = & \frac{\bar p'^4 + \bar p^4+ \bar k^4}
             {\bar p'^3 \, \bar p^3 \, \bar k^3} \,
        \Fg(\bar p', \bar p, \bar k) \, ,
\end{eqnarray}
\end{subequations}
where
\begin{equation}
\label{eq:integral}
  \F(\ptot,p, k) \equiv
  \frac{\ds \, \cs \, \alpha}{2(2 \pi)^3} \int \frac{d^2h}{(2 \pi)^2} \,
   2\h \cdot \Real \, \Fsol(\h;\ptot,p,k) .
\end{equation}
Here $\Fsol$ is the solution to the following integral equation
\begin{align}
\label{eq:lpmgen}
  2\h = &
  i \, \delta E(\h;\ptot,p,k)\, \Fsol(\h;\ptot,p,k)
\nonumber \\ &
  + g^2 \int \frac{d^2 q_\perp}{(2 \pi)^2} \,
    {\cal A}(q_\perp)
    \biggl\{
      \frac{1}{2}\ca \,
         \left[ \Fsol(\h;\ptot,p,k)-\Fsol(\h+\ptot \qperp;\ptot,p,k) \right]
\nonumber \\ & \hspace{9em}
      + (\cs - \frac{1}{2} \ca) \,
         \left[ \Fsol(\h;\ptot,p,k)-\Fsol(\h-k \, \qperp;\ptot,p,k) \right]
\nonumber \\ & \hspace{9em}
      + \frac{1}{2}\ca \,
          \left[ \Fsol(\h;\ptot,p,k)-\Fsol(\h-p \, \qperp;\ptot,p,k) \right]
     \biggr\} ,
\end{align}
which we will refer to as the LPM equation.  In this equation,
\begin {equation}
\label{eq:energy}
  \delta E(\h;\ptot,p,k) =
  \frac{\meffg}{2k}
  +\frac{\meffs}{2p}
  -\frac{\meffs}{2 \ptot}
  +\frac{h^2}{2pk \ptot}
\end{equation}
represents the energy denominator
$E_{g,\k}+E_{s,\p}-E_{s,\p'}$
in a $\ptot \leftrightarrow pk$ splitting/joining process.
Here, $m_{{\rm eff},s}$ is the $O(gT)$ effective thermal mass of hard
particles of species $s$.%
\footnote{
  We will not need explicit formulas for $m_{{\rm eff},s}$ in this
  paper.
}
The two-dimensional vector $\h$ is related to transverse
momentum and physically corresponds to the combination
\begin{equation}
\label{eq:h}
  \h = k {\bm p}_\bot - p {\bm k}_\bot .
\end{equation}
In terms of the qualitative discussion of Sec.\ \ref{sec:qualitative},
one can crudely think of the order of magnitude of $h$ as representing
$h \sim P \QT$.
The function ${\cal A}(\q_\perp)$ is the integrated correlator
\begin {equation}
  {\cal A}(\q_\perp) =
  \int \frac{d q_z}{2\pi} \,
  \bigl\langle\!\bigl\langle
      A^-(\omega,\q_\perp,q_z) [A^-(\omega,\q_\perp,q_z)]^*
  \bigr\rangle\!\bigr\rangle
  \bigg|_{\omega=q_z} ,
\label {eq:Adef}
\end {equation}
where $A^- \equiv A^0 - A^z$ and $\langle\!\langle A A^*
\rangle\!\rangle$ is the thermal Wightman gauge field correlator
(neglecting the momentum-conserving $\delta$-function).
If momentum transfers are small compared to typical plasma
particle momenta, then one may evaluate the self-energies in
this correlator in the hard thermal loop approximation.
For the case of equilibrium, ${\cal A}$ then has
the simple form \cite{Aurenche:2002pd}
\begin {equation}
   {\cal A}(q_\perp)
   = T \left( \frac{1}{q_\perp^2} - \frac{1}{q_\perp^2+\md^2} \right)
   = \frac{T \md^2}{q_\perp^2 (q_\perp^2 + \md^2)} .
\label {eq:A}
\end {equation}
As discussed in Ref.\ \cite{AMY2}, the last formula holds
more generally for the case of any
homogeneous plasma where the distribution of plasma particle momenta is
isotropic.  The ``temperature'' and Debye mass to use in
(\ref{eq:A}) in such a situation
are%
\footnote{
  For QCD with $\Nf$ flavors, and identical distributions of all quarks
  and anti-quarks, this would be
  $T_* = g^2
         \int \frac{d^3p}{(2\pi)^3} \, [ 6 f_g (1 + f_g) + 2\Nf f_q (1 - f_q) ]
         / \md^2$
  and
  $\md^2 = 2 g^2 \int \frac{d^3p}{(2\pi)^3} \, [ 6 f_g + 2\Nf f_q ] / p$.
}
\begin {equation}
   T = T_* \equiv
   \frac{
      \sum_s \bar\nu_s t_s \,
      \int\frac{d^3p}{(2\pi)^3} f_s(p) [1\pm f_s(p)]
   }{
      2 \sum_s \bar\nu_s t_s
      \int\frac{d^3p}{(2\pi)^3} \frac{f_s(p)}{p}
   }
   \,,
\end {equation}
\begin {equation}
   \md^2 =
     \textstyle{ 2 g^2
     \sum_s \bar\nu_s t_s \int\frac{d^3p}{(2\pi)^3} \frac{f_s(p)}{p} }
   \,,
\end {equation}
where $\bar\nu_s \equiv \nu_s/d_s$ is the number of degrees of freedom
of a species $s$ excluding color;
$f_s(p)$ is the phase space distribution of particles of type $s$
per spin and color degree of freedom; and the species sum is over gluons,
flavors of quarks, and flavors of anti-quarks.
For our analysis of splitting processes, this generalization to
isotropic non-equilibrium situations will only be valid if the
plasma particle distribution functions do not significantly change
over the course of a formation time.

The simple form (\ref{eq:A}) for ${\cal A}(q_\perp)$
is justified by our parametric
assumption $P \ll T/[\alphas^2 \ln(\alphas^{-1})]$.
We will give some discussion of what happens at higher $P$ in section
\ref{sec:run}, but we leave a full NLL calculation at higher
$P$ for future work.

In (\ref{eq:lpmgen}), the variable $\q_\perp$ represents the transverse
momentum exchange from individual $2{\rightarrow}2$ scattering processes.
Solving the integral equation for $\Fsol$ then accounts for summing up multiple
scattering into the LPM effect.  For a discussion with notation similar
to that used in this paper, see, for example, Ref. \cite{AMY1}.


\subsection {The leading log solution}
\label {sec:leadinglog}

The Coulomb logarithm arises in the LPM equation (\ref{eq:lpmgen})
from the region of $\q_\perp$ integration where $\md \ll q_\perp$
with $q_\perp$ still small enough that $\h+p'\q_\perp$ , $\h - k\q_\perp$
and/or $\h-p\q_\perp$ are still close to $\h$.
Roughly speaking, this is the
approximation that the important momentum transfers $\q_\perp$ from
individual collisions are large compared to $\md$ but small compared
to the total $Q_\perp$ from all the collisions during the formation
time.  Deep in the LPM regime, there will be more and more scatterings
in a formation time, and so $Q_\perp \gg q_\perp$.

In this same limit, the last term of (\ref{eq:energy}) will dominate,
and we can approximate
\begin{equation}
\label{eq:approxE}
  \delta E \simeq \frac{h^2}{2pk \ptot} \,.
\end{equation}
We can also approximate the differences of $\Fsol$'s on the
right-hand side of (\ref{eq:lpmgen}) by Taylor expansions,
keeping the first term which does not integrate to zero by
parity.  The result is
\begin {multline}
   2\h \simeq
   \frac{i h^2}{2pk\ptot} \Fsol(\h;\ptot,p,k)
   - \frac{g^2}{4}
   \biggl\{
      \frac{1}{2}\ca p'^2
      + (\cs - \frac{1}{2} \ca) k^2
      + \frac{1}{2}\ca p^2
   \biggr\}
\\ \times
   \nabla_{h}^2 \Fsol(\h;\ptot,p,k)
   \int \frac{d^2 q_\perp}{(2 \pi)^2} \,
   q_\perp^2 \, {\cal A}(q_\perp) .
\label {eq:diffeq1}
\end {multline}
From (\ref{eq:A}), the remaining $\q_\perp$ integral is logarithmically
UV divergent.  In the original integral, this divergence is cut off
when our approximation that $\h + p'\q_\perp$ etc.\ are close to $\h$ breaks
down.  Let $q_\perp \sim Q_{\perp0}$
represent any rough estimate of this breakdown scale.
Then, to leading order in logarithms,
\begin {equation}
   \int \frac{d^2 q_\perp}{(2 \pi)^2} \,
     q_\perp^2 \, {\cal A}(q_\perp)
   =
   T \md^2 \int \frac{d^2 q_\perp}{(2 \pi)^2} \,
     \frac{1}{q_\perp^2+\md^2}
   \simeq
   \frac{T \md^2}{4\pi} \, \ln(\QTo^2) ,
\label {eq:llog}
\end {equation}
where $\QTo \equiv Q_{\perp 0}/\md$.
The differential equation (\ref{eq:diffeq1}) for the
leading-log approximation to
$\Fsol(\h)$ can then be solved, applying the boundary conditions
that $\Fsol$ remain finite at $h=0$ and $h \to \infty$.
The result is%
\footnote{
  In solving (\ref{eq:diffeq1}), it is convenient to use rotational
  invariance to first write $\Fsol$ as $\h$ times a scalar function of
  $h^2$.
}
\begin {equation}
\label{eq:llsol}
  {\bm F}_{s0}(\h) =
  i \, 4p'pk
  \left[ \exp\left(-e^{\pm i\pi/4} \frac{h^2}{H_s^2}\right) -1 \right]
  \frac{\h}{h^2} \,.
\end{equation}
where
\begin{equation}
  H_s^2 =
  \left\{
     \frac{g^2}{2\pi} \mdsq T \, |p'pk|
     \left[
        \tfrac12\ca p'^2
        +(\cs -\tfrac{1}{2}\ca)k^2
        +\tfrac12\ca p^2
     \right]
     \ln(\QTo^2)
  \right\}^{1/2} .
\label {eq:H}
\end{equation}
The $\pm$ sign in (\ref{eq:llsol}) should be chosen as the sign
of $p'pk$ [which is negative in the case of (\ref{eq:gammagqq})],
but this sign will not have any effect on our final answers.
In terms of our previous qualitative discussion in
Sec.\ \ref{sec:qualitative} and qualitative identification
of $h \sim P Q_\perp$, the above squared width $H_s^2$
of the distribution (\ref{eq:llsol})
is parametrically of order
$H_s^2 \sim (P Q_\perp)^2 \sim P^2 \md^2 \QThat^2$ with $\QThat^2$ given
by (\ref{eq:Qperp1}).

Plugging (\ref{eq:llsol}) into Eq.\ (\ref{eq:integral}) for $\F$,
one finds the leading-log approximation
\begin {equation}
  \F \simeq
  \frac{d_s C_s \alpha}{(2\pi)^4} \, \sqrt{2} \, |p'pk| \, H_s^2 .
\end {equation}
Combining this with the equations (\ref{eq:gamma}) for the
splitting functions $\gamma$ gives the leading-log approximation
to our result, which just corresponds to replacing all three
logarithms on the right-hand side of (\ref{eq:muhat}) by
$\ln(\QTo^2)$:
\begin {align}
&
  \hat\mu_\perp^2(x_1,x_2,x_3;s_1,s_2,s_3)
  \simeq
  \frac{2 H_s^2}{\md^2 P^2}
\nonumber\\ & \qquad
  \simeq
  \frac{gT}{\md}
  \left[
    \frac{2}{\pi} \,
    x_1 x_2 x_3 \,
    \frac{P}{T}
  \right]^{1/2}
  \Biggl\{
    \Biggl[
       \tfrac12 (C_{s_2}+C_{s_3}-C_{s_1}) x_1^2
       + \tfrac12 (C_{s_3}+C_{s_1}-C_{s_2}) x_2^2
\nonumber\\ & \hspace{15em}
       + \tfrac12 (C_{s_1}+C_{s_2}-C_{s_3}) x_3^2
    \Biggr] \ln(\QTo^2)
  \Biggr\}^{1/2} .
\label {eq:muLL}
\end {align}

Our leading-log result is the same as that derived by other authors,%
\footnote{
  In particular, our leading-log result for $d\Gamma_{g{\to}gg}/dx$ using
  (\ref{eq:dGamma}), (\ref{eq:gamma_ggg}) and
  (\ref{eq:muLL}) is the same as Eqs. (19--20) of
  Ref.\ \cite{bottom_up} with their
  $\ln(\langle k_\perp^2\rangle/\md^2)$ identified as our
  $\ln(\QTo^2)$.
}
which were based on general formalisms for the LPM effect but applied
in a ``static'' approximation where the color fields of plasma particles
were treated as simply screened by a Debye mass in the same way as
static electric fields.
In this approximation, the correlator
${\cal A}(q_\perp)$ of (\ref{eq:A}) is replaced by%
\footnote{
  See, for example, Eq.\ (2.17) of Ref.\ \cite{BDMPS2}.
  The difference by an overall factor of $\pi T$ is because
  they have normalized their version $V(q_\perp^2)$ of
  ${\cal A}$ so that $\int d^2 q_\perp \, V(q_\perp^2) = 1$.
  See also Ref.\ \cite{djordjevic}.
}
\begin {equation}
  {\cal A}_{\rm static}(q_\perp) = \frac{T \md^2}{(q_\perp^2+\md^2)^2} .
\label {eq:Astatic}
\end {equation}
The difference between this static approximation and the actual case is
that the plasma screening of the non-static color electric and
magnetic fields generated by moving plasma charges is different from
the screening of the static, purely electric
fields.  However, the leading log result is generated by
$q_\perp \gg \md$, and in this case all screening effects can be
ignored.  Accordingly, (\ref{eq:Astatic}) is the same as
(\ref{eq:A}) when $q_\perp \gg \md$.
When we go to NLL order, individual momentum transfers $q_\perp$
of order $\md$ will be important, and we will find a difference between
a full treatment of plasma screening and the static approximation.


\subsection {The NLL solution}

To discuss the expansion, it is useful to introduce notation
\begin {equation}
   ({\bm f}_1, {\bm f}_2) \equiv
   \int\frac{d^2h}{(2\pi)^2} \> {\bm f}_1(\h) \cdot {\bm f}_2(\h)
\end {equation}
for the inner product of two vector functions of $\h$.
In this notation, the basic LPM equations (\ref{eq:integral}) and
(\ref{eq:lpmgen}) become
\begin {equation}
   {\cal F} =
   \frac{\ds \, \cs \, \alpha}{2(2 \pi)^3} \,
   \Real (\S,{\bm F}) ,
\end {equation}
and
\begin {equation}
   {\bm S} = C {\bm F} ,
\end {equation}
where we've defined the ``source'' ${\bm S}$ of the last equation by
\begin {equation}
   {\bm S} \equiv 2\h
\end {equation}
and $C$ is a linear operator defined such that $C {\bm F}$ is the
right-hand side of the LPM equation (\ref{eq:lpmgen}).

Let $C_0$ represent the approximation to $C$ which gives the leading-log
result.  Expanding to first order in powers of
$\delta C \equiv C-C_0$, we can write
\begin{align}
\label{eq:Fnew}
  (\S , {\bm F})
  &= (\S \, , \, C^{-1} \S)
\nonumber \\
  &= (\S \, , \, C_0^{-1} \S) - (\S, C_0^{-1} \delta C C_0^{-1} \S)
\nonumber \\
  &= 2(\S \, , \, C_0^{-1} \S) - (\S \, , \, C_0^{-1} C C_0^{-1} \S)
\nonumber \\
  &= 2(\S , {\bm F}_0) - ({\bm F}_0 \, , \, C {\bm F}_0) ,
\end{align}
and so
\begin {equation}
  \Real (\S , {\bm F})
  = 2 \Real (\S , {\bm F}_0) - \Real ({\bm F}_0 \, , \, C {\bm F}_0) .
\label{eq:SF}
\end {equation}
The first term is just twice the leading-log result,
which corresponds to
\begin {align}
  \Real (\S, {\bm F}_0) &= \frac{\sqrt2}{\pi} \, |p' p k| \, H^2
\nonumber\\
  &=
  \frac{g \md}{\pi} \, |p' p k|^{3/2}
  \left( \frac{T}{\pi} \right)^{1/2}
  \left\{
     \left[
        \tfrac12\ca p'^2
        +(\cs -\tfrac{1}{2}\ca)k^2
        +\tfrac12\ca p^2
     \right]
     \ln(\QTo^2)
  \right\}^{1/2} .
\label {eq:SF0}
\end {align}
We now need
to compute
\begin {equation}
  \Real ({\bm F}_0 \,,\, C {\bm F}_0)
  = \Real I_1
    + \tfrac12\ca \Real I_2(p'^2)
    + (\cs -\tfrac{1}{2}\ca) \Real I_2(k^2)
    + \tfrac12\ca \Real I_2(p^2) ,
\label {eq:FCF}
\end {equation}
where
\begin {equation}
  \Real I_1 \equiv
  \int \frac{d^2h}{(2\pi)^2} \>
  \Real[ {\bm F}_0(\h) \cdot i \, \delta E(h) \, {\bm F}_0(\h) ]
\end {equation}
and
\begin {equation}
  I_2(\kappa^2) \equiv
  g^2 \int \frac{d^2h}{(2\pi)^2} \> \frac{d^2 q_\perp}{(2 \pi)^2} \,
    {\cal A}(q_\perp) \,
    {\bm F}_0(\h) \cdot
    \left[ {\bm F}_0(\h)-{\bm F}_0(\h+\kappa \qperp) \right] .
\label {eq:I2def}
\end {equation}

In doing the $I_1$ integral, one can ignore the $m^2$ terms in
Eq. (\ref{eq:energy}) for $\delta E$ and use the approximation
(\ref{eq:approxE}).  The $m^2$ terms in (\ref{eq:energy})
are suppressed by order $m^2 P^2/h^2$ compared to the $h^2$
term.  In terms of the earlier qualitative discussion of
section \ref{sec:qualitative}, $h \sim P Q_\perp$, and so,
from (\ref{eq:Qperp2}), this
suppression factor is of order
\begin {equation}
   \frac{m^2 P^2}{h^2} \sim
   \frac{m^2}{Q_\perp^2} \sim \frac{1}{\QThat^2}
   \sim \left[\frac{P}{T} \ln\left(\frac{P}{T}\right) \right]^{-1/2} .
\label {eq:sorder}
\end {equation}
It is suppressed by a {\it power} of $P/T$, not simply a power of
the logarithm $\ln(P/T)$, and so these mass terms do not contribute at
any finite order in our inverse log expansion.  The $I_1$ integral
is trivial and then gives
\begin{equation}
\label{eq:I1}
  \Real \, I_1 = \frac{3|p'pk|H^2}{\pi\sqrt{2}}
  = \frac{3}{2} \, \Real(\S, {\bm F}_0) .
\end{equation}

The evaluation of $I_2$ is briefly outlined in the appendix and gives
\begin{equation}
\label{eq:I2}
  I_2(\kappa^2) =
  -\frac{g^2(p'pk)^2 T}{\pi^2}
  \int_0^{\infty} \frac{d\tau}{\tau}
  \left( 1-e^{-u_\kappa \tau}\right)
  \log\left(\frac{\tau+1}{\tau}\right) 
\end{equation}
where
\begin{equation}
\label{eq:s}
   u_\kappa \equiv e^{\pm i\pi /4}\, \frac{\md^2 \kappa^2}{2 H^2} \,.
\end{equation}
Parametrically, $u_\kappa$ is the same order as
(\ref{eq:sorder}) and so is small compared to one.
We therefore only need the small $u_\kappa$ expansion of the
integral (\ref{eq:I2}),%
\footnote
{
  The exact result for (\ref{eq:I2}) can be expressed (somewhat
  uselessly) in terms of the
  Meijer $G$ function as
  $I_2(\kappa^2) =
     -\frac{g^2(p'pk)^2 T}{\pi^2} \left[
     \frac{\pi^2}{4}+\frac{1}{2}\,(\gammaE+\ln u_\kappa)^2
     -G^{3,1}_{2,3} \left( u_\kappa \left|
        \substack{0 \,,\, 1 \\ 0 \,,\, 0 \,,\, 0} \right. \right)
     \right]
  $.
}
which is derived in the appendix:
\begin{equation}
  I_2(\kappa^2) =
  - \frac{g^2(pk\ptot )^2 T}{\pi^2} \,
  ( 2-\gammaE-\ln u_\kappa ) u_\kappa +\O(u_\kappa^2) ,
\label {eq:I2expand}
\end{equation}
and so
\begin{equation}
\label{eq:I2smalls}
  \Real \, I_2(\kappa^2) =
  - \frac{g^2(pk\ptot )^2 T}{\pi^2 \sqrt{2}}
  \left( 2-\gammaE +\frac{\pi}{4}-\ln|u_\kappa| \right) |u_\kappa|
  + \O(u_\kappa^2) .
\end{equation}
Using (\ref{eq:H}) and (\ref{eq:SF0}), this can be rewritten as
\begin{equation}
  \Real \, I_2(\kappa^2) =
  - \frac{1}{2} \, \Real(\S, {\bm F}_0) \,
  \frac{\kappa^2 \, \ln\left(\frac{2 \xi H_s^2}{\md^2 \kappa^2}\right)}
       { \left[
           \tfrac12\ca p'^2
           +(\cs -\tfrac{1}{2}\ca)k^2
           +\tfrac12\ca p^2
         \right]
         \ln(\QTo^2) } ,
\end {equation}
where $\xi$ is the NLL constant under the log defined in (\ref{eq:xi}).
Putting our results for $I_1$ and $I_2$ into
(\ref{eq:SF}) and (\ref{eq:FCF}), the NLL result for $\Real(\S,{\bm F})$ is
\begin {align}
  & \Real({\bm S},{\bm F}_s)
  =
  \Real({\bm S},{\bm F}_0) \,
\nonumber\\ & \qquad \times
  \frac12
  \left\{
    1 +
    \frac{
       \tfrac12\ca p'^2 \ln\left(\frac{2 \xi H_s^2}{\md^2 p'^2}\right)
       +(\cs -\tfrac{1}{2}\ca)k^2 \ln\left(\frac{2 \xi H_s^2}{\md^2 k^2}\right)
       +\tfrac12\ca p^2 \ln\left(\frac{2 \xi H_s^2}{\md^2 p^2}\right)
    }{
       \tfrac12\ca p'^2 \ln(\QTo^2)
       +(\cs -\tfrac{1}{2}\ca)k^2 \ln(\QTo^2)
       +\tfrac12\ca p^2 \ln(\QTo^2)
    }
  \right\}
\label {eq:NLLblah}
\end {align}
with $H_s$ given by (\ref{eq:H}).
At leading-log order, where all logarithms are treated as the same,
the multiplicative correction above simply reduces to a factor of one.
At NLL order, its effect on the leading-log result (\ref{eq:SF0})
is simply to replace the curly brackets in (\ref{eq:SF0}) by the
numerator of the big fraction in (\ref{eq:NLLblah}).
This corresponds
to the iterated NLL solution (\ref{eq:mu2}) quoted in
section \ref{sec:result},
when the splitting functions $\gamma$ are
written in the form of (\ref{eq:gammas}).
Note that $H_s^2$ in (\ref{eq:NLLblah})
depends on $\QTo$.  As mentioned earlier, the self-consistent equation
(\ref{eq:muhat}) provides an answer which is equivalent at NLL order but
which does not depend on specifying an initial guess $\QTo$.


\subsection {Dynamic vs.\ static treatment of screening}

Since the static approximation (\ref{eq:Astatic}) to screening is sometimes
used in the literature, we will take a moment to discuss how the
NLL order result would have been different had we made that
approximation.  This is simple to do by comparing with (\ref{eq:A})
and noting that
\begin {equation}
   {\cal A}_{\rm static}
   = \md^2 \frac{\partial {\cal A}}{\partial(\md^2)}
   = |u_\kappa| \frac{\partial {\cal A}}{\partial |u_\kappa|} \,,
\label {eq:Atrick}
\end {equation}
where the last equality uses (\ref{eq:s}) considering $H$ as fixed.
The only change comes in the calculation of $I_2$.
Applying (\ref{eq:Atrick}) to (\ref{eq:I2smalls}),
\begin {equation}
  \Real \, I_2(\kappa^2) \rightarrow
  \Real \, I_2^{\rm static}(\kappa^2) =
  - \frac{g^2(pk\ptot )^2 T}{\pi^2 \sqrt{2}}
  \left( 1-\gammaE +\frac{\pi}{4}-\ln|u_\kappa| \right) |u_\kappa|
  + \O(u_\kappa^2) .
\end{equation}
This produces the same NLL result (\ref{eq:muhat})
as the fully dynamic case but with
\begin {equation}
  \xi \to \xi_{\rm static}
  \equiv \exp\left(1 - \gammaE + \tfrac{\pi}{4}\right) 
  \simeq 3.3474 \,.
\end {equation}


\section {Numerical Results}
\label{sec:numerics}

\begin{figure}
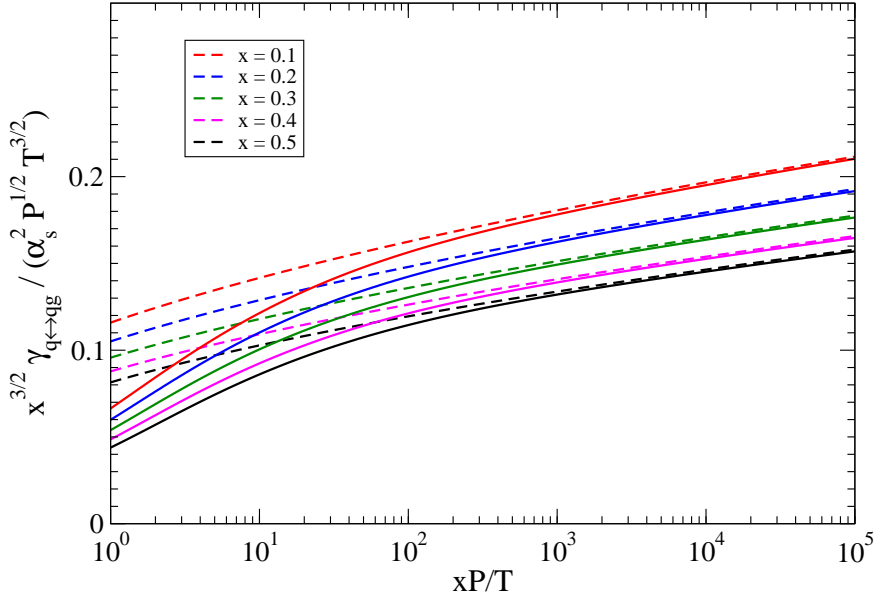

  \centerbox{0.7}{ququg_Nf3_p_rescale1.eps}
  \caption{\label{fig:qgq1}
  Similar to Fig.\ \ref{fig:ggg_rescale} but for
  $\gamma_{q \leftrightarrow gq}$.
  $x$ is the momentum fraction of the gluon.
}
\end{figure}

\begin{figure}
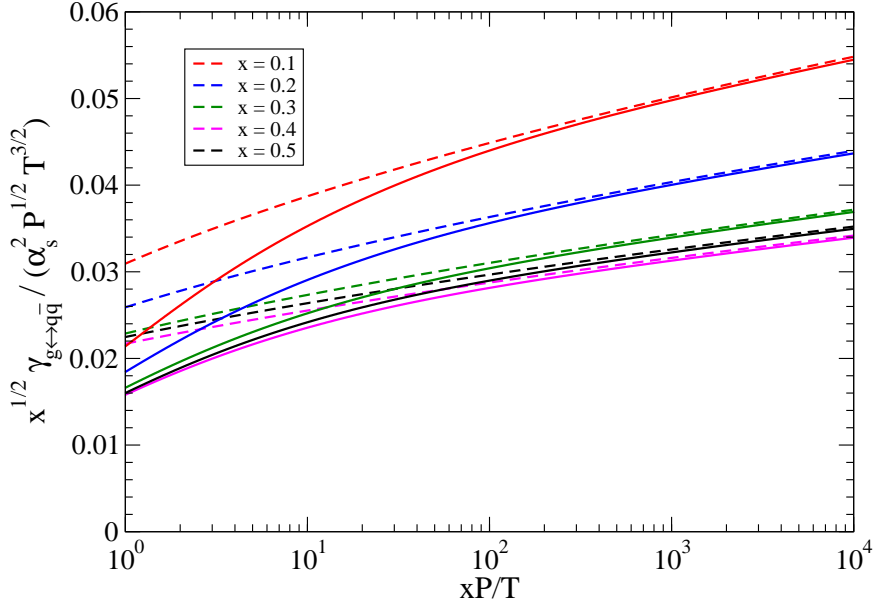

  \centerbox{0.7}{gququbar_Nf3_p_rescale.eps}
  \caption{\label{fig:gqq}
  Similar to Fig.\ \ref{fig:ggg_rescale} but for
  $\gamma_{g \leftrightarrow q\bar q}$.
}
\end{figure}

In the introduction, we have already discussed for $g\to gg$
the comparison of
NLL results with a full computation at leading order in powers of
$\alphas$.
The scaling in Fig.\ \ref{fig:ggg_rescale} was chosen by inspection
of the small $x$ limit of the NLL result given by (\ref{eq:gammas})
and (\ref{eq:muhat}), choosing powers of $x$ so that the
curves will be similar for small enough $x$.
We give similar results for $\gamma_{q\to qg}$ and
$\gamma_{g \to q\bar q}$ in Figs.\ \ref{fig:qgq1} and \ref{fig:gqq}.
The fact that the small $x$ curves in these figures do not fall
closer together is because we have not quite gone to small
enough $x$ and because we have not bothered with factors of
$\ln(1/x)$ in our consideration of how to scale the axis.

\begin{figure}
  \centerbox{0.7}{ququg_Nf3_p_rescale2.eps}
  \caption{\label{fig:qgq2}
  The case $q \to g q$ for $x > 0.5$.
}
\end{figure}

For the case of $q\to gq$, there is no final state symmetry that
relates results for $x$ to $1{-}x$, so we show $x > 0.5$ results for
this case in Fig.\ \ref{fig:qgq2}, scaled so that the curves are
similar in the $1{-}x \to 0$ limit.

From these graphs, one can confirm for all splitting processes
the general claim made in the introduction that the deviation of
the NLL approximation from the full result at leading order in
powers of $\alphas$ is roughly 20\% or better for $p_< \gtrsim 10 T$.


\section {The case of large \boldmath$\alphas\ln(P/T)$}
\label{sec:run}

Throughout, we have assumed that $\alphas$ is small enough that we
can ignore all effects suppressed by powers of $\alphas$.
In particular, we have ignored the issue of what renormalization scale
$\alphas$ should be evaluated at: $\md$, $T$, $Q_\perp$, or $P$?
Formally,
\begin {equation}
   \alphas(\mu')
   = \alphas(\mu)
     \left[1 + \beta_0 \alphas \ln\left(\frac{\mu'^2}{\mu^2}\right)
             + \O(\alphas^2 \ln^2) \right] ,
\end {equation}
and so we can ignore the ambiguity of scale choice if
$\alphas \ln(\mu'/\mu)$ is small for relevant different possibilities
$\mu$ and $\mu'$.

In this paper, we have studied the deep LPM regime, where
$\ln(Q_\perp/\md) \sim \ln(P/T)$ is large.
Nonetheless, we have so far implicitly assumed that
$\alphas \ln(Q_\perp/\md) \sim \alphas \ln(P/T)$ is small,
since we have ignored all higher order corrections in $\alphas$,
including the running of the coupling constant.
This assumption is also implicit in the full leading-order
results of previous work, such as the LPM equation
(\ref{eq:lpmgen}).  However, in applications there could
be a significant difference between $\alphas(\md)$ and
$\alphas(Q_\perp)$ --- that is, $\alphas \ln(Q_\perp/\md)$ may
be large. In this section, we discuss what to
do in that situation.

Earlier, we made the parametric assumption
$P \ll T/[\alphas^2 \ln(\alphas^{-1})]$ in order to simplify
the analysis of this paper, which is a first step towards
computing the NLL result for the general case.  With this
assumption, the expansion
parameter $\alphas \ln(P/T)$ is never large in the weak
coupling limit, since then
$\alphas \ln(P/T) \lesssim \alphas \ln(\alphas^{-2}) \ll 1$.
In general, different choices of scale $\mu$ in $\alphas(\mu)$
are formally not significant in the weak coupling limit if
the different scales only differ by some power of the
coupling.  So, for example, there is no significant
difference between $\alphas(\md)$ and $\alphas(T)$ in the
weak coupling limit.

In this section, we will abandon the restriction that
$P \ll T/[\alphas^2 \ln(\alphas^{-1})]$ and consider extremely
large $P$ for which $\alphas(Q_\perp)$ may be significantly different
than $\alphas(\md) \simeq \alphas(T)$.
We will correspondingly
restrict ourself to a leading-log analysis, though now resumming
the large factors of $\alphas \ln(Q_\perp/\md)$ due to running
of the coupling constant.


\subsection{Large \boldmath$q_\perp$ behavior of \boldmath${\cal A}(q_\perp)$}
\label{sec:largeA}

In our previous analysis, the leading-log result was dominated by
$\md \ll q_\perp$, where the gauge field correlation function had the
simple form
\begin {equation}
  {\cal A}(q_\perp) \simeq \frac{T \md^2}{q_\perp^4}
  \qquad
  \mbox{($\md \ll q_\perp \ll T$)}.
\label {eq:Aapprox1}
\end {equation}
This can be understood as arising from the lower part of the scattering
diagram of Fig.\ \ref{fig:diagram}.  The $1/q_\perp^4$ is the
contribution $(-\omega^2+q^2)^{-2}$ of the gauge propagator to the
rate, with $\omega$ set equal to $q^z$ as in the general definition
(\ref{eq:Adef}) of ${\cal A}(q_\perp)$.  Setting $\omega=q^z$ is
a reflection of energy conservation for the high-energy particle
(the top solid line of Fig.\ \ref{fig:diagram}) in the limit that $Q \ll P$.
The factor of $T \md^2$ in (\ref{eq:Aapprox1})
reproduces the contribution of the plasma particle and
its phase space integral, integrated over $q_z$ as in (\ref{eq:Adef}).
It's appearance is more transparent if we write an explicit formula
for the Debye mass,
\begin {equation}
  T \md^2 =
  g^2 \sum_s \nu_s t_{R_s}
  \int \frac{d^3\ell}{(2\pi)^3} \> f_s(\ell) \, [1 \pm f_s(\ell)] .
\end {equation}
Here one can see the $g^2$ associated with the contribution
to the rate of the
bottom vertex in Fig.\ \ref{fig:diagram}, the Bose
or Fermi distribution $f(\ell)$ for the probability of
encountering a plasma particle with momentum $\ell$, a final
state Bose enhancement or Fermi blocking factor
$1 \pm f(|\bell+\q|) \simeq 1 \pm f(\ell)$, a sum over species
$s$ of the plasma particle, and the appropriate group factors and
numbers of degrees of freedom.

For the opposite limit of $q_\perp \gg T$, one obtains a similar
expression, but there is no final state factor $1\pm f$.  That's because
if $\ell \sim T$ is a plasma particle momentum, and $q_\perp \gg T$,
then $|\bell+\q| \gg T$ and so $1 \pm f(\bell+\q) \simeq 1$.
The form of ${\cal A}(\q_\perp)$ in this limit is
\begin {equation}
  {\cal A}(q_\perp) \simeq \frac{g^2 {\cal N}}{q_\perp^4}
  \qquad
  \mbox{($q_\perp \gg T$)} ,
\label {eq:Aapprox2}
\end {equation}
where
\begin {equation}
   {\cal N} \equiv
  \sum_s \nu_s t_{R_s}
  \int \frac{d^3\ell}{(2\pi)^3} \>  f_s(\ell)
\end {equation}
is a measurement of the density of plasma particles, weighted by
group factors.
For QCD with $\Nf$ massless fermion flavors, ${\cal N}$ is
\begin {equation}
   {\cal N} = \frac{\zeta(3)}{\zeta(2)}
              \Bigl( \ta + \tfrac32 \Nf \tf \Bigr)
              \tfrac13 T^3
   = \frac{\zeta(3)}{\zeta(2)} \left( 1 + \tfrac14 \Nf \right) T^3 ,
\label {eq:N}
\end {equation}
where $\zeta(s)$ is the Riemann zeta function.
Compare this formula to (\ref{eq:md}) to see the difference between
$T \md^2$ in (\ref{eq:Aapprox1}) and
$g^2 {\cal N}$ in (\ref{eq:Aapprox2}).

For 3-flavor QCD, $g^2 {\cal N}$ is about 15\% smaller than
$\md^2 T$.  For any phenomenologically relevant values of the
coupling $\alphas$, this difference is unlikely to be significant
compared to other corrections, such as higher-order effects.
So, one could reasonably just start from our earlier results based on
the original formula (\ref{eq:A}) for ${\cal A}(T)$.
However, the conceit of this paper is to work out precise results
in the formal limit of arbitrarily weak coupling (and large
logarithms), and so for sufficiently high energy jets we
use (\ref{eq:Aapprox2}).

Now consider our previous leading-log analysis.  The Coulomb logarithm was
generated from (\ref{eq:llog}):
\begin {equation}
   \int \frac{d^2 q_\perp}{(2 \pi)^2} \,
     q_\perp^2 \, {\cal A}(q_\perp)
   \simeq
   \int_{\sim\md}^{\sim Q_\perp} \frac{q_\perp \> d q_\perp}{2\pi} \,
     \frac{T \md^2}{q_\perp^2}
   \simeq
   \frac{T \md^2}{2\pi} \, \ln\left(\frac{Q_\perp}{\md}\right)
   \qquad
   (Q_\perp \ll T).
\end {equation}
For $Q_\perp \gg T$, there are instead two logarithmic contributions,
coming from the different integration regions represented
by (\ref{eq:Aapprox1}) and (\ref{eq:Aapprox2}):
\begin {align}
   \int \frac{d^2 q_\perp}{(2 \pi)^2} \,
     q_\perp^2 \, {\cal A}(q_\perp)
   &\simeq
   \frac{T \md^2}{2\pi} \, \ln\left(\frac{T}{\md}\right)
   + \frac{g^2 {\cal N}}{2\pi} \, \ln\left(\frac{Q_\perp}{T}\right)
   \qquad
   (Q_\perp \gg T)
\nonumber\\
   &\simeq
   \frac{(T \md^2-g^2{\cal N})}{2\pi} \, \ln\left(\frac{T}{\md}\right)
   + \frac{g^2 {\cal N}}{2\pi} \, \ln\left(\frac{Q_\perp}{\md}\right)
   .
\end {align}
If $P \gg T/[\alphas^4 \ln(\alphas^{-1})]$,
so that $Q_\perp/T \gg T/\md$ by (\ref{eq:QT}), then the second logarithm
dominates.  Since in this section we will be interested in $P$ large
enough that there is significant running of the coupling constant
(and so formally $P$ larger than $T$ times any fixed power of
$1/\alphas$),
we shall henceforth assume this is the case.  If we ignored the
running of the coupling, we'd have
\begin {equation}
   \int \frac{d^2 q_\perp}{(2 \pi)^2} \,
     q_\perp^2 \, {\cal A}(q_\perp)
   \simeq
   \int_{\sim \md}^{\sim Q_\perp} \frac{q_\perp \> d q_\perp}{2\pi} \,
     \frac{g^2 {\cal N}}{q_\perp^2}
   \simeq
   \frac{g^2 {\cal N}}{2\pi} \, \ln\left(\frac{Q_\perp}{\md}\right)
   \qquad
   \bigl( Q_\perp \gg \frac{T}{\alphas^4 \ln(\alphas^{-1})} \bigr)
\label {eq:llogbig}
\end {equation}
in this case.
Given our assumpations and approximations, we could have just as
well made the lower limit $T$ instead of $\md$ in (\ref{eq:llogbig}).
However, because of the small difference between
$g^2{\cal N}$ and $\md^2 T$, taking the lower limit to be
$\md$ is probably slightly better in practice.


\subsection{Running of coupling with $q_\perp$}
\label {sec:run1}

We begin with a discussion of the scale of the coupling $g$ in the
$g^2 {\cal A}(q_\perp; g^2)$ factor in the LPM equation
(\ref{eq:lpmgen}),
where we now write ${\cal A}(q_\perp; g^2)$ instead of ${\cal A}(q_\perp)$
to emphasize the fact that ${\cal A}$ depends on $g^2$, e.g.\ as in
(\ref{eq:Aapprox2}).
Physically, this factor is proportional to the rate for
a high-energy particle to scatter off of a
plasma particle, with momentum transfer $q_\perp$.
(See, for example, the discussion in Ref.\ \cite{AMY1}.)
A momentum transfer of $q_\perp$ corresponds
to an impact parameter between the high-energy particle and the
plasma particle of order $1/q_\perp$.  If this distance is
very, very small, then $g^2$ in this Coulomb scattering amplitude
should be correspondingly smaller because of the anti-screening of
the QCD vacuum.  (For $1/q_\perp \ll$ the Debye screening length
$1/\md$, medium effects are ignorable, and so the relevant consideration
for such a collision is of the running coupling constant in vacuum.)
The upshot is that the appropriate scale for $g^2$ in this
scattering rate should be set by $q_\perp$.  This is precisely
what we would get by summing up all 1-loop bubbles on the
exchanged gluon line.  We therefore expect that the LPM equation
(\ref{eq:lpmgen}) should be modified by replacing $g^2$ by
$g^2(q_\perp)$ so that it becomes
\begin{equation}
  2\h =
  i \, \delta E(\h;\ptot,p,k)\, \Fsol(\h;\ptot,p,k)
  + \int \frac{d^2 q_\perp}{(2 \pi)^2} \,
    g^2(q_\perp) \, {\cal A}(q_\perp; g^2(q_\perp))
    \Bigl\{ \cdots \Bigr\} .
\label {eq:lpmrun}
\end{equation}
This is the same prescription as used in a related context
by Peshier in an analysis of collisional energy loss \cite{peshier}.

Now repeat the leading-log analysis we reviewed in
Section \ref{sec:leadinglog}.
Correspondingly, we'll use the 1-loop renormalization
group result for $g^2(q_\perp)$,
\begin {equation}
   g^2(\mu) = \frac{1}{-\bar\beta_0 \ln(\mu^2/\Lambda^2)}.
\label {eq:grun}
\end {equation}
For QCD,
$\Lambda$ represents $\Lambda_{\rm QCD}$,
and
\begin {equation}
  \bar\beta_0 =
  - \frac{(11 \ca - 4 \Nf \tf)}{48 \pi^2}
  = - \frac{(33 - 2 \Nf)}{48 \pi^2}
  < 0 .
\end {equation}
(For QED, $\Lambda$ represents the ultraviolet Landau pole,
and $\bar\beta_0 > 0$.)
The integral analogous to $g^2$ times
(\ref{eq:llogbig}) is
\begin {align}
   \int \frac{d^2 q_\perp}{(2 \pi)^2} \,
     g^2(q_\perp) \, q_\perp^2 \, {\cal A}(q_\perp; g^2(q_\perp))
   &\simeq
   \int_{\sim \md}^{\sim Q_\perp} \frac{q_\perp \> d q_\perp}{2\pi} \,
     \frac{g^4(q_\perp) {\cal N}}{q_\perp^2}
\nonumber\\
   &\simeq
   \frac{{\cal N}}{8\pi\bar\beta_0^2}
   \int_{\sim \md}^{\sim Q_\perp}
   \frac{d q_\perp}{q_\perp \ln^2(q_\perp/\Lambda)}
\nonumber\\
   &=
   \frac{{\cal N}}{8\pi \bar\beta_0^2}
   \left[
      \frac{1}{\ln(\md/\Lambda)} - \frac{1}{\ln(Q_\perp/\Lambda)}
   \right] .
\label{eq:runint}
\end {align}
Following through the previous leading-log derivation, we will
get the same result for $F_0$ and $H$ but with the replacement
\begin {equation}
  g^2 \md^2 T \ln(\QTo^2)
  \to
  \frac{{\cal N}}{2 \bar\beta_0^2}
   \left[
      \frac{1}{\ln(\md/\Lambda)} - \frac{1}{\ln(Q_{\perp0}/\Lambda)}
   \right]
  =
  {\cal N} \,
  \frac{[g^2(\md) - g^2(Q_{\perp0})]}{-\bar\beta_0}
\label {eq:logsub}
\end {equation}
in (\ref{eq:H}).  The correspondence with the original leading-log
result is easier to see if we use (\ref{eq:grun}) to rewrite this
in the equivalent form
\begin {equation}
  g^2 \md^2 T \ln(\QTo^2)
  \to
  g^2(\md) \, g^2(Q_\perp) \, {\cal N}\,
  \ln(\QTo^2)
  .
\label {eq:yuri}
\end {equation}

It is interesting to note that, because of asymptotic freedom in QCD,
the running-coupling formula on the right-hand side of (\ref{eq:logsub})
is finite as $P$ (and so $Q_\perp$) becomes infinite.  In this limit,
the value of $\alphas(Q_\perp)$ is irrelevant---a fact slightly obscured
by (\ref{eq:yuri}) but made clear by (\ref{eq:logsub}).  In this case,
the momentum transfers $q_\perp$ in individual $2{\to}2$ scatterings
which dominate the integral (\ref{eq:runint}) range, roughly speaking,
from $T$ up to those where $\alphas(q_\perp)$ can first be considered
small compared to $\alphas(T)$.  In particular, once $Q_\perp$ is
large enough, the scale of $q_\perp$ does not continue to grow as
one increases $P$ and $Q_\perp$.


\subsection{Remaining \boldmath$g^2$ and synthesis}

There remains one other factor of $g^2$ in the problem, which is
the cost for emitting the energetic, bremsstrahlung particle.
Because of the LPM effect, a bremsstrahlung gluon cannot resolve
individual $2{\to}2$ collisions with the plasma
but is only sensitive to the net deflection of the
high energy particles over the entire formation time.
The relevant scale for the $g^2$ cost of bremsstrahlung is
therefore plausibly the total momentum transfer%
\footnote{
  $1/Q_\perp$ is the transverse
  distance corresponding to the deflection of the high-energy particle
  due to $2{\to}2$ scatterings during a formation time.
  To see this, consider that if the particle picks up transverse
  momentum $Q_\perp$ in that time, then it's transverse velocity will
  be of order $Q_\perp/P$.  Multiplying this transverse velocity
  by a formation time
  $t_{\rm form} \sim P/Q_\perp^2$ (\ref{eq:tform}) then gives
  $1/Q_\perp$.  The same scale is also the scale of the quantum
  mechanical uncertainty of the transverse position during a formation
  time (which is why LPM interference can occur).  In contrast,
  the scale $1/q_\perp$ relevant to section \ref{sec:run1}
  is the typical transverse distance between
  the high energy particle and the plasma particles it is scattering
  from.
}
(\ref{eq:Qperp2}) $Q_\perp \sim \md (P/T)^{1/4}\ln^{1/4}(P/T)$.
[We will ignore the issue of whether this last $P$
should be $P$ or $xP$ or
$(1-x)P$ but will assume that $x$ and $1-x$ are large enough that
there is not much difference in $\alphas$.]
We therefore propose that the correct leading-log formula
in the case of small $\alphas$ but large
$\alphas \ln(P/T)$ is given by (i) replacing
$\alphas$ by $\alphas(Q_\perp) \simeq \alphas(\md \QTo)$
in (\ref{eq:gammas}), and then (ii)
using (\ref{eq:logsub}) to replace (\ref{eq:muLL}) by
\begin {align}
  \mu_\perp^2 \equiv
  \md^2 \hat\mu_\perp^2 &\simeq
  (T {\cal N})^{1/2}
  \left[
    \frac{2}{\pi} \,
    x_1 x_2 x_3 \,
    \frac{P}{T}
  \right]^{1/2}
\nonumber\\ & \times
  \biggl[
       \tfrac12 (C_{s_2}+C_{s_3}-C_{s_1}) x_1^2
       + \tfrac12 (C_{s_3}+C_{s_1}-C_{s_2}) x_2^2
       + \tfrac12 (C_{s_1}+C_{s_2}-C_{s_3}) x_3^2
  \biggr]^{1/2}
\nonumber\\ & \times
  \biggl[
    \frac{g^2(\md)-g^2(Q_{\perp 0})}{-\bar\beta_0}
  \biggr]^{1/2} .
\end {align}
[Note that it is the combination $\mu_\perp^2 = \md^2 \hat\mu_\perp^2$
which appears in (\ref{eq:gammas}).]
Here, $Q_{\perp 0}$ is chosen with order of magnitude given by
(\ref{eq:Qperp2}), or one could simply self-consistently choose
$Q_{\perp 0} = \mu_\perp$.

For the case of large $g^2\ln(\QTo^2)$, there is not an
obvious benefit to pushing further to find the NLL solution to
the running LPM equation (\ref{eq:lpmrun}).  For large
$g^2\ln(\QTo^2)$, the
expansion parameter $[\ln(Q_\perp/\md)]^{-1}$ is parametrically
the same order as $\alphas$, and so one cannot obviously justify
neglecting corrections that are higher order in $\alphas$ that
have not been included in the running of the coupling.
On the other hand, it would be nice to have a single, well-defined
formula that interpolated our previous NLL result into the
realm of large $g^2\ln(\QTo^2)$.  But we shall not pursue this
here.

We have made plausibility arguments about which renormalization scales
should be used for evaluating the coupling $g^2$.  It's possible,
however, that if one starts looking at corrections suppressed by
$\alphas$ that there are other large logarithms which arise, unrelated
to the runnings we have described.  To settle the issue definitely,
it would be nice to have explicit weak-coupling
calculations beyond leading order in $\alphas$ of
a well-defined physical quantity that is dominated by
particle splitting processes.


\begin{acknowledgments}

We are indebted to Guy Moore who, after reading the original version of
this manuscript, pointed out to us that our analysis of the NLL result
breaks down for $p \gtrsim T/[\alphas^2\ln(\alphas^{-1})]$,
and that the same effect required
corrections to our analysis of Sec.\ \ref{sec:run}.
We are also indebted to the referee, who led us to
realize another error in our original analysis of Sec.\ \ref{sec:run}
concerning our treatment of the running coupling.
We also thank Yuri Kovchegov for useful and interesting conversations,
and in particular for pointing out to us that the Coulomb logarithm for
the case of running coupling constants can be written in the form of
(\ref{eq:yuri}).
This work was supported, in part, by the U.S. Department
of Energy under Grant No.~DE-FG02-97ER41027.

\end{acknowledgments}


\appendix

\section{\boldmath$I_2$}

\subsection {The \boldmath$I_2$ integral}

Here is one way to do the integral $I_2(\kappa^2)$ of (\ref{eq:I2def}).
First use the standard trick of rewriting
\begin {equation}
   \frac{1}{q^2+m^2} =
   \int_0^\infty d\lambda \> e^{-\lambda(q^2+m^2)}
   ,
\end {equation}
so that
\begin {equation}
  I_2(\kappa^2) =
  g^2 T \int_0^\infty d\lambda
  \int \frac{d^2h}{(2\pi)^2} \> \frac{d^2 q_\perp}{(2 \pi)^2} \,
  \left[ e^{-\lambda q_\perp^2} - e^{-\lambda (q_\perp^2+\md^2)} \right]
    {\bm F}_0(\h) \cdot
    \left[ {\bm F}_0(\h)-{\bm F}_0(\h+\kappa \qperp) \right] .
\end {equation}
Next rewrite (\ref{eq:llsol}) in the form
\begin {equation}
   {\bm F}_0(\h) = C \left( e^{-A h^2} - e^{-\epsilon h^2} \right)
              \frac{\h}{h^2} ,
\end {equation}
where $C = i4p'pk$, $A=e^{\pm i\pi/4} / H^2$, and the limit
$\epsilon \to 0^+$ is taken at the end of the day.
By introducing $\epsilon$, we can split $I_2$ into
\begin {equation}
  I_2(\kappa^2) = {\cal I}(0) - {\cal I}(\kappa^2) ,
\label {eq:I2calI}
\end {equation}
where
\begin {equation}
  {\cal I}(\kappa^2) =
  g^2 T \int_0^\infty d\lambda \> (1-e^{-\lambda \md^2})
  \int \frac{d^2h}{(2\pi)^2} \> \frac{d^2 j}{(2 \pi)^2 \kappa^2} \,
    e^{-\lambda j^2/\kappa^2}
    {\bm F}_0(\h) \cdot
    {\bm F}_0(\h+\j)
\label{eq:calI}
\end {equation}
and we have introduced the notation $\j\equiv -\kappa \q_\perp$.
The integral ${\cal I}$ would be divergent if we had not
introduced some sort of regulator like the $\epsilon$.
Now note that (\ref{eq:calI}) has the form of a convolution,
and so by Fourier transformation we can recast it as a single
two-dimensional integral over a Fourier conjugate variable which
we'll call $\B$.
Two-dimensional Fourier
transformation of $\h$ or $\j$ takes
\begin {align}
  {\bm F}_0(\h) &\to
  -\frac{i C \B}{2\pi B^2} \left( e^{-B^2/4A} - e^{-B^2/4\epsilon} \right)
  ,
\\
  e^{-\lambda j^2/\kappa^2}  &\to
  \frac{\kappa^2}{4\pi\lambda} \, e^{-\kappa^2 B^2/4\lambda} ,
\end {align}
and so the integral becomes
\begin {equation}
  {\cal I}(\kappa^2) =
  \frac{g^2 C^2 T}{16\pi^3} \int_0^\infty \frac{d\lambda}{\lambda}
  (1-e^{-\lambda \md^2})
  \int \frac{d^2B}{B^2} \>
  e^{-\kappa^2 B^2/4\lambda}
  \left( e^{-B^2/4A} - e^{-B^2/4\epsilon} \right)^2
  .
\end {equation}
By (\ref{eq:I2calI}), then
\begin {equation}
  I_2(\kappa^2) =
  \frac{g^2 C^2 T}{16\pi^3} \int_0^\infty \frac{d\lambda}{\lambda}
  (1-e^{-\lambda \md^2})
  \int \frac{d^2B}{B^2} \>
  (1 - e^{-\kappa^2 B^2/4\lambda})
  \left( e^{-B^2/4A} - e^{-B^2/4\epsilon} \right)^2
  .
\end {equation}
We can now safely take the limit $\epsilon \to 0$, giving
\begin {equation}
  I_2(\kappa^2) =
  \frac{g^2 C^2 T}{16\pi^3} \int_0^\infty \frac{d\lambda}{\lambda}
  (1-e^{-\lambda \md^2})
  \int \frac{d^2B}{B^2} \>
  (1 - e^{-\kappa^2 B^2/4\lambda})
  e^{-B^2/2A}
  .
\end {equation}
Finally, doing the $\B$ integral and changing variables
from $\lambda$ to $\tau \equiv 2\lambda/A \kappa^2$ gives (\ref{eq:I2}).


\subsection {Small \boldmath$u_\kappa$ expansion}

To expand (\ref{eq:I2}) in powers of $u_\kappa$, it is convenient to
notice that $I_2(0) = 0$ and then instead expand
\begin {align}
   \frac{\partial I_2}{\partial u_\kappa}
   &=
   -\frac{g^2(p'pk)^2 T}{\pi^2}
   \int_0^{\infty} d\tau \>
   e^{-u_\kappa \tau}
   \log\left(\frac{\tau+1}{\tau}\right)
\nonumber\\
   &=
   -\frac{g^2(p'pk)^2 T}{\pi^2}
   \,
   \frac{[\gammaE + \ln u_\kappa - e^{u_\kappa} \operatorname{Ei}(-u_\kappa)]}
        {u_\kappa} \,,
\end {align}
where Ei is the exponential integral.  The small $u_\kappa$ expansion
is
\begin {equation}
   \frac{\partial I_2}{\partial u_\kappa}
   =
   -\frac{g^2(p'pk)^2 T}{\pi^2}
   \left( 1 - \gammaE - \ln u_\kappa \right).
\end {equation}
Integrating both sides and using $I_2(0)=0$ then yields
(\ref{eq:I2expand}).


\end{document}